\documentclass[aps,prl,superscriptaddress,longbibliography,twocolumn]{revtex4-2}

\usepackage{graphicx}
\usepackage{amsmath}
\usepackage{amssymb}
\usepackage{hyperref}
\usepackage[utf8]{inputenc}
\usepackage{mathtools}
\usepackage[english]{babel}
\usepackage{bbm}
\hypersetup{colorlinks=true, linkcolor=blue, citecolor=blue, urlcolor=blue}
\usepackage{xcolor}
\usepackage{braket}
\usepackage{bbold}
\usepackage[normalem]{ulem}

\newcommand{\ie}{i.\,e.,\ }

\newcommand{\eg}{e.\,g.,\ }

\newcommand{\bea}{\begin{eqnarray}}
\newcommand{\eea}{\end{eqnarray}}

\usepackage{pgf}
\usepackage{layouts}

\begin{document}
\title{Deterministic generation of photonic entangled states using decoherence-free subspaces}

\author{Oriol Rubies-Bigorda}
\email{orubies@mit.edu}
\affiliation{Physics Department, Massachusetts Institute of Technology, Cambridge, Massachusetts 02139, USA}
\affiliation{Department of Physics, Harvard University, Cambridge, Massachusetts 02138, USA}
\author{Stuart J. Masson}
\affiliation{Department of Physics, Columbia University, New York, New York 10027, USA}
\author{Susanne F. Yelin}
\affiliation{Department of Physics, Harvard University, Cambridge, Massachusetts 02138, USA}
\author{Ana Asenjo-Garcia}
\affiliation{Department of Physics, Columbia University, New York, New York 10027, USA}

\begin{abstract}
We propose the use of collective states of matter as a resource for the deterministic generation of quantum states of light, which are fundamental for quantum information technologies. Our minimal model consists of three emitters coupled to a half-waveguide, i.e., a one-dimensional waveguide terminated by a mirror. Photon-mediated interactions between the emitters result in the emergence of bright and dark states. The dark states form a decoherence-free subspace, protected from dissipation. Local driving of the emitters and control of their resonance frequencies allows to perform arbitrary quantum gates within the decoherence-free subspace. Coupling to bright states facilitates photon emission, thereby enabling the realization of quantum gates between light and matter. We demonstrate that sequential application of these gates leads to the generation of photonic entangled states, such as Greenberger-Horne-Zeilinger and one- and two-dimensional cluster states.
\end{abstract}
\maketitle

Large entangled states are a crucial resource in quantum technologies~\cite{QuantumComputation,QuantumEntangl,QuantumInfo,QuantumMetrology,QuantumCommu}, yet they are typically hard to create. A prominent example of multipartite entanglement is provided by Greenberger–Horne–Zeilinger (GHZ) states~\cite{GHZ_1,GHZ_2}, useful for quantum metrology~\cite{metrology_intro}, where qubits exist in a superposition of either all in one state or all in the other. Another relevant class of entangled states are cluster states~\cite{Briegel_MeasurementBAsedQC, Briegl_2001,Briegel_2009}, where qubits are arranged in a lattice and entangled with their nearest neighbors. Two-dimensional (2D) cluster states are particularly significant, as they constitute a universal resource for measurement-based quantum computation~\cite{Briegel_MeasurementBAsedQC,Briegel_2009,Briegel_MBQC}.

Deterministic photon entanglement can be achieved by interfacing light with multilevel quantum emitters~\cite{KimbluDuanGate,SWAP_gate}. Over the past decades, various protocols have been developed for sequential photon emission, including coupling an emitter to cavities~\cite{Cirac_cavity} or one-dimensional photonic channels~\cite{Lindner_cluster} to generate 1D cluster states. This has been extended to 2D cluster states using time-delayed feedback in chiral waveguides~\cite{Pichler_cluster}. Other approaches involve arrays of emitters coupled to multiple waveguides~\cite{Lindner_2D,multilevel_QD,Wallraff_2d_cluster} or an ancillary atom interacting with free-space atomic arrays~\cite{Bekenstein_Rydberg,Malz_Rydberg}. Following these proposals, recent experiments have exploited the multi-level structure of a single emitter to realized few-photon GHZ and cluster states using quantum dots~\cite{QuantumDot}, superconducting qubits~\cite{Wallraff_states,Painter_states,Wallraff_2d_cluster}, and neutral atoms~\cite{Rempe2022}.

\begin{figure}[hbt!]
    \includegraphics[width=0.95\columnwidth]{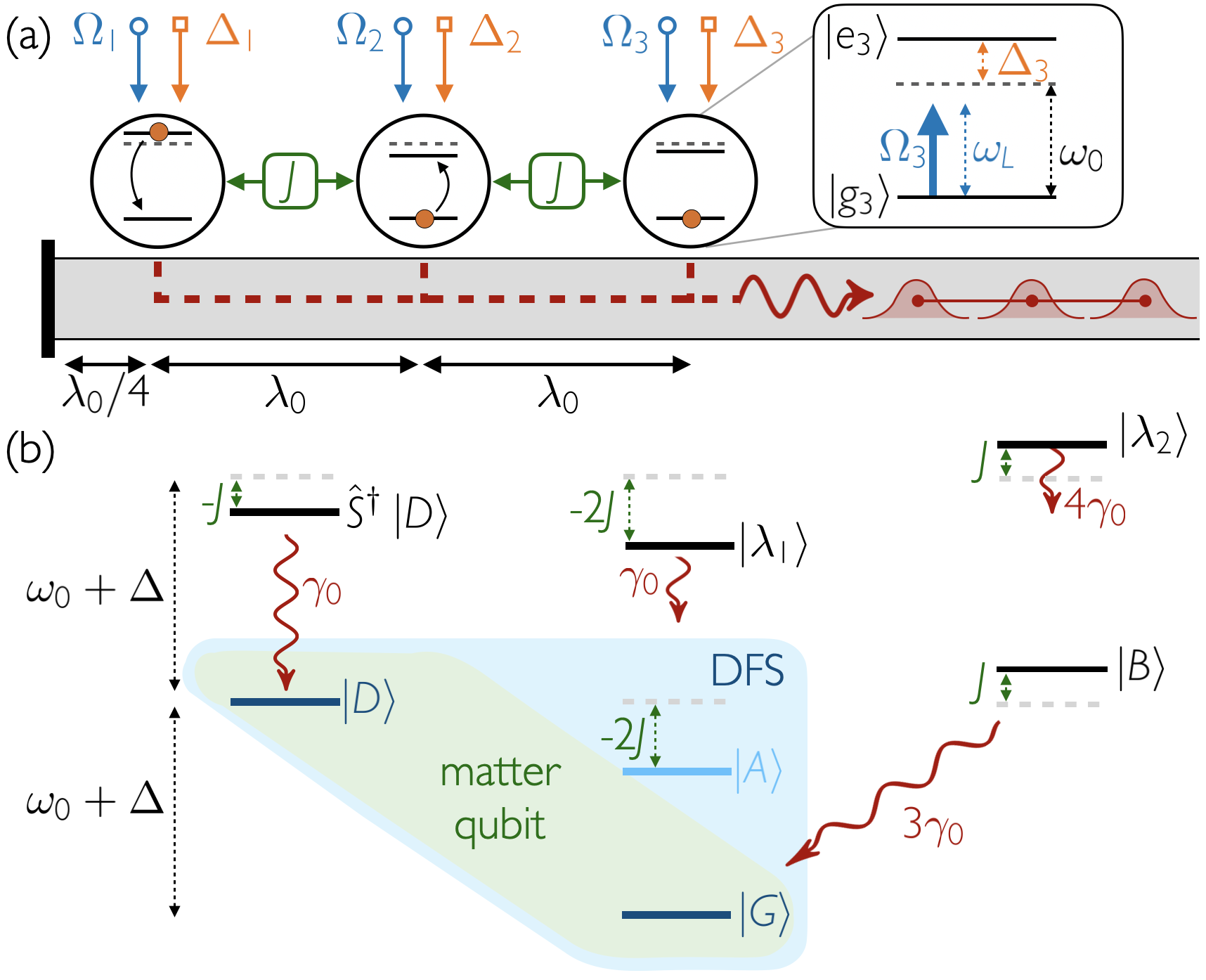}
    \caption{Local operations together with long-range interactions between three emitters coupled to a half-waveguide result in the generation of entangled photons. (a) Proposed setup:  external fields allow for tunable and local frequency shifts $\Delta_{1,3} = \Delta$ and $\Delta_2 = \Delta + \delta$ and Rabi frequencies $\Omega_n$ (orange and blue control lines, respectively). Coherent exchange interactions $J$ between neighboring emitters (depicted in green) can be engineered via external couplers. (b) Collective basis states in the absence of drive and for $\delta = -J$. The ground state $\ket{G}$ and the two single-excitation states $\ket{D}$ and $\ket{A}$ are dark and form a decoherence-free subspace (DFS). $\ket{G}$ and $\ket{D}$ form the matter qubit that remains entangled with the emitted photons, whereas $\ket{A}$ is used as an auxiliary state for conditional photon emission. The remaining states are bright (decay is shown in red).}
    \label{fig: sketch}
\end{figure}

Multiple emitters can interact through the electromagnetic vacuum, leading to modified radiative properties~\cite{Dicke,Goban_Waveguide,Ana_subradiance,Kirchmair,Goban_lightmatter,Ephi_mirror,Painter_QO,Bloch_mirror,Subradiance_storage_control}. This can create a decoherence-free subspace (DFS)\cite{Zanardi_DFS,Facchi_DFS,Lidar_DFS,Beige_DFS} of dark states decoupled from the environment and thus from dissipation. Prior theoretical work has shown that multiple ``collective'' qubits can be encoded in this subspace, enabling universal quantum computation via slow gates within the DFS~\cite{Lidar_DFS_QC,Beige_DFS_QC,Alejandro_DFS}. Interactions between emitters also result in the emergence of ``bright'' states, which allow to emit few-photon pulses~\cite{Alejandro_PhotonGeneration1,Alejandro_PhotonGeneration2}.

In this Letter, we propose generating entangled photonic states by harnessing the collective states of a few two-level emitters. A minimal configuration of three emitters coupled to a waveguide terminated by a mirror produces a qutrit in the decoherence-free subspace. We demonstrate a full set of fast logical quantum gates for the qutrit by combining local frequency shifts and driving fields together with controlled exchange interactions between pairs of emitters. Local frequency shifts further result in controlled coupling between collective dark and bright states, which enables the design of entangling quantum gates between the matter qutrit and photonic qubits with arbitrary temporal profiles. Sequential application of these quantum gates on the emitters results in the generation of photonic entangled states. We exemplify the procedure by theoretically demonstrating the preparation of GHZ, 1D and 2D cluster states.

\emph{System}.--- Our minimal setup consists of three two-level emitters~\footnote{While the minimal setup described in the main text consists of three emitters, entangled photonic states can also be attained with only two emitters coupled to the half-waveguide. However, this requires a more complex control of the system, as discussed in the Supplemental Material.} with transition frequency $\omega_0$ coupled to a one-dimensional waveguide terminated by a mirror, i.e., a half-waveguide [see Fig.~\ref{fig: sketch}(a)]. Emitters are located at distances $x_n = (n+1/4) \lambda_0$ from the mirror, where $\lambda_0=2\pi c/\omega_0$. The half-waveguide ensures light is emitted into one direction without requiring chirality~\cite{WillOliver_unidirectionalemission} or the recombination of both pulses~\cite{Cirac_recombination} to enforce a single output channel.

When the photon propagation time between emitters is much shorter than their characteristic timescale, the waveguide photons can be traced out using the Born-Markov approximation~\cite{fayard}. This yields an effective master equation for the emitters' density matrix, $\dot{\hat{\rho}} = -i [ \hat{H}, \hat{\rho}] + \mathcal{L}[\hat{\rho}]$, where $\hbar=1$. For the specific locations of the emitters along the half-waveguide, the waveguide-mediated coherent interaction is zero and the Lindbladian reduces to (see Supplemental Material [SM]~\footnote{See the Supplemental Material for a detailed derivation of the effective master equation of the emitters, the matter gates, the emission of single photons with arbitrary wavepackets and the light-matter gates, as well as their respective fidelities and robustness to imperfections. The Supplemental Material also describes an alternative protocol based on two emitters only, and includes the additional Refs.~\cite{CohenTanoudgi_book,Meystre_book,fidelity_3,fidelity_1,fidelity_2,Photon_wavepacket,Photon_wavepacket2}\label{footnote: SM}}\newcounter{firstfootnote}
\setcounter{firstfootnote}{\value{footnote}})
\begin{equation}
\label{eq: Lindbladian}
    \mathcal{L}[\hat{\rho}] = 3 \gamma_0 \left( \hat{S} \hat{\rho} \hat{S}^\dagger - \frac{1}{2} \{ \hat{S}^\dagger \hat{S}, \hat{\rho}\} \right),
\end{equation}
where $\gamma_0$ denotes the single-emitter decay rate. Photon emission is governed by the collective jump operator $\hat{S} = \sum_{n=1}^{3} \hat{\sigma}_n /\sqrt{3}$, where $\hat{\sigma}_n = |g_n \rangle \langle e_n|$ are the individual lowering operators, and occurs at a rate three times that of a single emitter. 

The system’s coherent dynamics are governed by the Hamiltonian
\begin{align}
\label{eq: Hamiltonian_main}
    \hat{H} &= \Delta\sum_n \hat{\sigma}_n^\dagger \hat{\sigma}_n + \delta \hat{\sigma}_2^\dagger \hat{\sigma}_2 + J \left( \hat{\sigma}_1^\dagger \hat{\sigma}_2 + \hat{\sigma}_2^\dagger \hat{\sigma}_3 + h.c \right) \nonumber \\
    &+ \sum_n \left( \Omega_n e^{-i (\omega_L - \omega_0) t} \hat{\sigma}_n^\dagger + h.c. \right).
\end{align}
Here, the first and third emitters are shifted by $\Delta$ from the bare transition frequency $\omega_0$, whereas the second emitter acquires a shift $\Delta + \delta$ [orange control lines in Fig.~\ref{fig: sketch}(a)]. Additionally, neighboring emitters undergo coherent exchange interactions at a rate $J$. 
Finally, fields external to the half-waveguide drive the emitters at frequency $\omega_L$ with arbitrary position-dependent Rabi frequencies $\Omega_n$~\cite{Kirchmair}. All Hamiltonian parameters can be varied over time.

Collective emission gives rise to states that are decoupled from dissipation. This decoherence-free subspace~\cite{Beige_DFS,Alejandro_DFS} is spanned by the ground state $\ket{G} \equiv |ggg\rangle$ and the single-excitation collective states $\ket{D} = (\ket{egg} - \ket{gge}) / \sqrt{2}$ and $\ket{A} = (\ket{egg} - 2\ket{geg} + \ket{gge}) / \sqrt{6}$. The remaining single-excitation state $\ket{B} = (\ket{egg} + \ket{geg} + \ket{gge}) / \sqrt{3}$ is bright and decays to the ground state at a rate $3 \gamma_0$. The three two-excitation states are bright, with decay rates ranging from $\gamma_0$ to $4\gamma_0$. For $\delta = -J$ and no drive (\ie $\Omega_n = 0$), the ground and single-excitation states are eigenstates of the Hamiltonian in Eq.~(\ref{eq: Hamiltonian_main}), each acquiring a frequency shift~[see Fig.~\ref{fig: sketch}(b) and SM]. Detunings $\delta \neq -J$ induce coherent coupling between $\ket{A}$ and $\ket{B}$, while $\ket{D}$ remains unaffected.

\begin{figure*}
    \includegraphics[width=\textwidth]{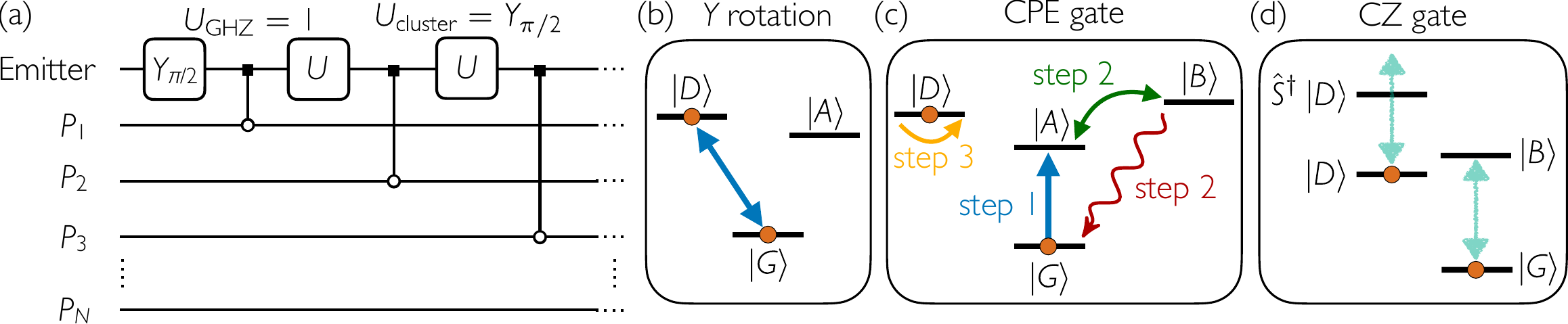}
    \caption{Gate protocol for the generation of entangled photon states. (a) Quantum circuit schematic. The matter qubit and photonic qubits are initialized in the ground state. The emission of each entangled photonic qubit is achieved through the sequential operation of a light-matter CPE gate (\ie emission of a photon conditional on the state of the matter qubit) followed by a unitary operation $U$ on the matter qubit. To generate GHZ states, $U\equiv\mathcal{I}$, and to generate one-dimensional cluster states, $U\equiv Y_{\pi/2}$. (b-d) Implementation of quantum gates in the collective basis. (b) $Y_{\pi/2}$ rotation of the matter qubit is attained by driving the $\ket{D} \leftrightarrow \ket{G}$ transition. Population of higher excited states is suppressed by the large off-resonance shift $J$ and decay rate $\gamma_0$ of the bright two-excitation states. (c) The CPE gate is attained in three steps. First, the amplitude in $\ket{G}$ is transferred to the auxiliary dark state $\ket{A}$. Then, a nonzero detuning $\delta \neq -J$ is applied to the second emitter, coupling $\ket{A}$ and $\ket{B}$ and leading to photon emission. Finally, a phase gate is applied to compensate the phase acquired by $\ket{D}$ during the first two steps. (d) The CZ gate is realized by scattering a photonic qubit from the collective system. For large coherent coupling $J$, the photon is resonant with $\ket{G} \leftrightarrow \ket{B}$ but far off-resonant with $\ket{D} \leftrightarrow \hat{S}^\dagger \ket{D}$, and consequently acquires different phases depending on the state of matter qubit.}
    \label{fig: circuit_gates}
\end{figure*}

Prior proposals to prepare time-binned entangled photon states use a single multi-level emitter with two long-lived states forming a matter qubit, from which conditional photon emission (\ie a CPE gate~\footnote{While the conditional photon emission (CPE) gate is sometimes referred to as a CNOT gate in literature, the CPE gate only implements the CNOT gate provided that the input photon is in the vacuum or ground state. }) is possible~\cite{Lindner_cluster,Pichler_cluster,Painter_states,Wallraff_states}. In contrast, our approach leverages the collective states of a three-emitter system. We encode the matter qubit in the states $\ket{D}$ and $\ket{G}$. Coupling to the auxiliary dark state $\ket{A}$ and bright states enables controlled light-matter interactions, and allows to implement entangling light-matter CPE and CZ gates. Combined with full control over the emitters' quantum state within the DFS, one can generate entangled photon states by sequentially applying matter and light-matter gates, as detailed below.

\emph{Matter gates in the DFS}.--- Arbitrary quantum operations on the DFS-encoded qutrit state $\ket{\psi_{DFS}}=d(t) e^{-i \omega_0 t} \ket{D} + g(t) \ket{G} + a(t) e^{-i \omega_0 t} \ket{A}$ can be realized via rotations between two pairs of states $\{\ket{G}, \ket{D}, \ket{A}\}$ together with independent control over their phases~\cite{qutrit_1,qutrit_2,qutrit_3}. The implementation of high-fidelity quantum gates requires that control sequences do not populate states outside the DFS. This is naturally the case in the absence of drive and provided that $\delta = -J$ [see Fig.~\ref{fig: sketch}(b)]. Driving can induce transitions to bright collective states in higher excitation manifolds. However, when the decay rates of the bright states are much larger than the drive strength (\ie $\Omega \ll \gamma_0$), these states are only virtually populated. In this regime, the system dynamics are effectively confined to the DFS via the quantum Zeno effect. This constraint on the Rabi frequency limits gate speeds. The use of external couplers with $J \gg \Omega$ further suppresses excitation via photon blockade, thereby enabling stronger driving and faster gate times, $T \sim \Omega^{-1} \gg 1/\sqrt{\gamma_0^2 + J^2}$.

Arbitrary rotations between $\ket{G}$ and $\ket{D}$ require the collective system to be driven resonantly ($\omega_L = \omega_0$ and $\Delta=0$) with the profile of $\ket{D}$, \ie $\Omega_1 = -\Omega_3 = \Omega e^{i \phi} / \sqrt{2}$ and $\Omega_2=0$ [see Fig.~\ref{fig: circuit_gates}(b)]. Error arises from residual coupling of $\ket{D}$ to the bright states $S^\dagger \ket{A}$ and $S^\dagger \ket{B}$, as well as coupling of $\ket{A}$ to $S^\dagger \ket{D}$; all these transitions are off-resonant by at least $J$. For $\Omega^2 \ll \gamma_0^2 + J^2$, evolution over a time $T = \theta / \Omega$ implements, up to small corrections, the quantum gate (see SM)
%the system dynamics of the system after a time $T=\theta / \Omega$ implements up to small corrections the quantum gate (see SM)
\begin{equation}
\label{eq:rotation_main}
    \mathcal{R}_{DG}(\theta,\phi,\chi) = \begin{pmatrix}
\cos(\theta) & - i e^{i \phi} \sin (\theta) & 0\\
- i e^{-i \phi} \sin (\theta) & \cos(\theta) & 0 \\
0 & 0 & e^{ i \chi}
\end{pmatrix},
\end{equation}
in the basis $\{ \ket{D}, \ket{G}, \ket{A} \}$ and for $\chi = 2 J T$. Notably, $ \mathcal{R}_{DG} (\pi /4,-\pi/2,\chi) \equiv \mathrm{Y}_{\pi/2}$ corresponds to the $\pi/2$-rotation around the $y$ axis of the Bloch sphere of the matter qubit. Similarly, arbitrary rotations between the dark states $\ket{G}$ and $\ket{A}$ are obtained by matching the drive profile to the profile of $\ket{A}$, $\Omega_1 = - \Omega_2/2 = \Omega_3 = \Omega e^{i\phi}/\sqrt{6} $.
The phase acquired by $\ket{D}$ during the operation is $\chi = -2 J T$ (see SM).  

Arbitrary phase control requires three phase gates (see SM for details). First, coherent coupling in the absence of drives (\ie $J = - \delta \neq 0$ and $\Omega_n=\Delta=0$) gives rise to a phase for state $\ket{A}$ of $-2 J T$. Conversely, applying only an equal detuning to all qubits (\ie $\Delta \neq 0$ and $\Omega_n = J = \delta = 0$) results in equal phases acquired by $\ket{D}$ and $\ket{A}$ of $\phi = \Delta T$. Control over the phase of $|G\rangle$ requires drive of the form $\Omega_n = \Omega/\sqrt{3}$, which couples off-resonantly every state in the DFS to a higher-excited state, introducing different Stark shifts to each state. Since the detuning $\Delta$, the drive detuning $\omega_0 - \omega_L$, and the coherent external coupling $J$ can typically be much larger than $\gamma_0$, the phase gates can be performed in a time $T \ll \gamma_0^{-1}$.

\emph{Light-matter gates}.--- Generation of entangled photonic states requires entangling gates between the DFS and the photonic qubits. A conditional photon emission (CPE) gate can be implemented in three steps, as illustrated in Fig.~\ref{fig: circuit_gates}(c). First, we transfer the amplitude in the ground state $\ket{G}$ to the auxiliary dark state $\ket{A}$, \ie $d_0 \ket{D} + g_0 \ket{G} \rightarrow d_0 e^{i \chi} \ket{D} - i g_0 \ket{A}$. Then, a non-zero detuning $\delta \neq -J$ is applied to the second emitter, which couples $\ket{A}$ and $\ket{B}$. This leads to the emission of a photon in the $k$-th step only if the system is in $\ket{A}$, resulting in the state $d_0 e^{i \chi'} \ket{D} \otimes \ket{0_k} + g_0 \ket{G} \otimes \ket{1_k} $. Finally, the phase $\chi'$ acquired by $\ket{D}$ during both operations is compensated by applying a phase gate. Arbitrary temporal photon wavepackets are attained by appropriately controlling $\delta (t)$ (see SM).

A CZ gate is implemented by interfacing a previously-created photonic qubit with the system of emitters. More precisely, the incoming photon couples the states $\ket{G}$ and $\ket{B}$, as well as $\ket{D}$ and $\hat{S}^\dagger\ket{D}$. Notably, the frequencies associated to both transitions differ by $2J$, resulting in the photon acquiring different phases depending on the state of the matter qubit [see Fig.~\ref{fig: circuit_gates}(d)]. A photon with a small bandwidth (\ie a large temporal width) resonant with the $\ket{G} \leftrightarrow \ket{B}$ transition will acquire a phase flip when scattering from $\ket{G}$. If the matter qubit is in $\ket{D}$, the photon is detuned from the $ |D \rangle \leftrightarrow \hat{S}^\dagger | D \rangle$ transition by $2 J$ and acquires a phase $\phi_D = \pi + 2 \arctan(4J/\gamma_0)$ (see SM). In the regime of strong coherent external coupling, $J \gg \gamma_0$, the far off-resonant photon acquires a vanishing phase, $\phi_D \rightarrow 0$, and we implement a CZ gate that only flips the sign of the state with one photon and the matter qubit in $\ket{G}$.

\begin{figure*}
    \includegraphics[width=\textwidth]{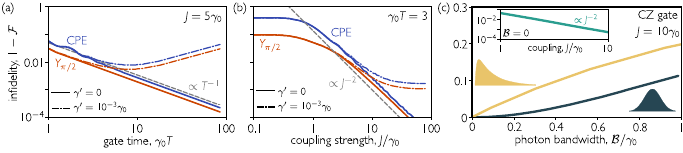}
    \caption{Gate infidelity $\epsilon = 1 - \mathcal{F}$. (a,b) Infidelity of the CPE gate (blue) and the $Y_{\pi/2}$ rotation (orange) as a function of (a) gate time $T$ and (b) strength of the coherent coupling $J$. Without emission into undesired channels, $\gamma ' =0$, the infidelity scales as $\epsilon \propto \gamma_0 J^{-2} T^{-1}$ for $J \gg \gamma_0$ (see grey dashed lines as a reference). For $\gamma ' \neq 0$, the infidelity increases for long gate times and saturates for large coherent couplings. (c) Infidelity of the CZ gate as a function of the photon bandwidth $\mathcal{B}$ at $J=10 \gamma_0$ for a Gaussian photon (dark green) and a photon obtained via constant coupling between $\ket{A}$ and $\ket{B}$ during emission (yellow). The inset shows the infidelity of the CZ gate for an ideal photon with zero bandwidth as a function of the coupling strength $J$.}
    \label{fig: errors}
\end{figure*}

Finally, the collective level structure also enables to easily disentangle the emitters from the photonic states by slightly modifying the CPE gate for the last photonic qubit. In particular, one needs to apply a $\pi$-rotation that transfers $\ket{D}$ to $\ket{G}$, $Y_\pi \equiv \mathcal{R}_{DG}(\pi/2,-\pi/2,\chi) $, between the transfer $\ket{G} \rightarrow \ket{A}$ and the photon emission process. Then, the collective system finishes the protocol in the ground state $\ket{G}$ while the last photonic qubit is still emitted in an entangled fashion.

\emph{Generating GHZ and cluster states}.--- Sequential application of rotations on the matter qubit and the light-matter CPE gate results in a train of entangled photonic qubits in different time bins [see Fig.~\ref{fig: circuit_gates}(a)]. An $m$-qubit photonic GHZ state is generated by applying the CPE gate $m$ times on the initial state $(\ket{G}+\ket{D}) / \sqrt{2}$, with matter-photon disentanglement performed in the final step.

Similarly, an $m$-qubit 1D cluster state is obtained by applying a $\pi/2$-rotation on the matter qubit ($\mathrm{Y}_{\pi/2}$) followed by a CPE gate $m$ times each. Following the proposal in Ref.~\cite{Pichler_cluster}, higher-dimensional entanglement structures can be generated by letting the $k$-th photonic qubit interact with the system of emitters a second time between the emission of photon $k+N-1$ and $k+N$. A two-dimensional photonic cluster state in an $M \times N$ lattice (where $MN$ is the total number of emitted photons) is obtained if the matter qubit applies the controlled phase gate $CZ$ on the $k$-th photon. For that, the emitted photons are reflected back to the system of emitters by an additional switchable mirror (\eg an additional emitter) placed at the transmitting end of the half-waveguide~\cite{Painter_states}. The transmission line needs to be long enough such that it can support $N$ temporally non-overlapping photons.

\emph{Timescales and errors}.--- Gates accumulate errors arising from (i) small populations of the bright states outside the DFS, (ii) independent photon emission at a rate $\gamma '$ into modes different than those of the half-waveguide and described by the Lindbladian $\sum_{n=1}^3 \gamma ' (\hat{\sigma}_n \hat{\rho} \hat{\sigma}_n^\dagger - \{\hat{\sigma}_n^\dagger \hat{\sigma}_n \}/2)$, (iii) imperfect control of the Hamiltonian, or (iv) fluctuations in the ideal positions of the emitters (see SM). The error of the $Y_{\pi/2}$ gate and of the transfer from $\ket{G}$ to $\ket{A}$ in the CPE gate is analytically (see SM) and numerically [Fig.~\ref{fig: errors}(a,b)] found to scale as $\epsilon \propto \gamma_0 T^{-1} (J^2+\gamma_0^2)^{-1}$. Optimal gate fidelities are achieved for large coherent couplings $J$ and long gate times $T$ (\ie weak drives  $\Omega$), which better confine dynamics to the DFS by enhancing the Zeno effect and photon blockade. In the presence of emission into undesired channels, the error eventually increases for large $T$, thereby setting an optimal gate time. 

The infidelity of the CZ gate for an ideal photonic qubit with zero bandwidth scales as $\propto \gamma_0^2/ J^{2}$ due to its finite detuning from the $ |D \rangle \leftrightarrow \hat{S}^\dagger | D \rangle$ transition, as shown in the inset of Fig.~\ref{fig: errors}(c) and demonstrated in the SM. For photonic wavepackets with finite bandwidth $\mathcal{B}$, different frequency components acquire slightly different phases during the scattering process associated to the CZ gate. This leads to a reduction of the fidelity with increasing bandwidth, as shown in Fig.~\ref{fig: errors}(c). For photons obtained by setting a constant coupling between $\ket{A}$ and $\ket{B}$ during the photon emission process, the infidelity for small bandwidths scales as $\mathcal{B} / \gamma_0$. Shaping the wavepacket can reduce the infidelity, which scales as $\mathcal{B}^2 / \gamma_0^2$ for Gaussian wavepackets.

\emph{Implementation}.--- Our protocols for photonic entangled-state generation can be implemented with superconducting qubits~\cite{Painter_QO,WillOliver_unidirectionalemission,Painter_states,Kirchmair,Wallraff_2d_cluster} coupled to a transmission line with an open end, which effectively acts as a mirror~\cite{SC_mirror}. Local frequency shifts and driving fields with arbitrary phase profiles can be achieved through external control lines~\cite{Kirchmair}, and strong time-dependent coherent interactions between different qubits can be implemented via capacitive couplers~\cite{WillOliver_unidirectionalemission,Kirchmair}. While transmon qubits are not strictly two-level systems, they  can exhibit anharmonicities $\gtrsim 10^2\gamma_0$~\cite{WillOliver_unidirectionalemission}, such that their multi-level structure can be neglected. Finally, they typically exhibit dissipative couplings to the waveguide that are orders of magnitude larger than their dephasing or non-radiative decay rates, a crucial requirement to generate high-fidelity entangled states with a large number of photonic qubits. Alternative implementations include atoms or quantum dots coupled to nanophotonic  waveguides~\cite{Goban_lightmatter,Goban_Waveguide,Collective_QD} or cavities. In this case, dark states could be directly addressed using free-space driving fields, with coherent exchange coupling engineered via Rydberg interactions~\cite{Rydberg_Misha}.

\emph{Conclusions and outlook}.--- Collective states of a few two-level emitters coupled to a half-waveguide can be harnessed to generate photonic entangled states. Our protocol relies on the sequential application of fast gates between the system’s dark states and the conditional emission of photons, which serves as an entangling light-matter operation. Fast gate operations between the dark states are enabled by coherent exchange interactions between emitters, while conditional photon emission is achieved by selectively coupling dark and bright states via local frequency shifts. The protocol’s duration is ultimately limited by photon emission times, which could be further reduced by leveraging superradiance in larger emitter arrays.

Our protocol could be extended to realize other entanglement structures, such as tensor network states~\cite{Cirac_tensor_networks,Malz_tensornetworks}. Alternatively, systems with four or more qubits support multi-excitation dark states~\cite{ours_PRA_waveguide}, which could enable the generation of higher-dimensional photonic states~\cite{high_dimensional_states_1,high_dimensional_states_2} using qudits instead of qubits. Beyond generating quantum states of light, collective dark states and their selective coupling to radiating states can serve as quantum memories~\cite{Ana_subradiance,Subradiance_storage_control,Raphael_waveguide}, facilitating communication and networking between distant nodes~\cite{Networks_Cirac,Kimble2008,Oliver_communication}.

\emph{Acknowledgments}.--- O.R.-B. acknowledges support from Fundación Mauricio y Carlota Botton and from Fundació Bancaria “la Caixa” (LCF/BQ/AA18/11680093). We acknowledge additional support by the National Science Foundation through the CAREER Award (No. 2047380) and the QII-TAQS program (No. 1936345), the Air Force Office of Scientific Research through their Young Investigator Prize (No. 21RT0751), as well as by the David and Lucile Packard Foundation. S.F.Y. acknowledges funding by the NSF through the CUA PFC (PHY-2317134) and through PHY-2207972.

\let\oldaddcontentsline\addcontentsline% Store \addcontentsline
\renewcommand{\addcontentsline}[3]{}% Make \addcontentsline a no-op
%\bibliographystylesupp{apsrev4-2}
\bibliography{reference_file}
\vspace{.5cm}
%\printbibliography[heading=none]
\let\addcontentsline\oldaddcontentsline% Restore \addcontentsline

\clearpage
\pagebreak

%%%%%%%%%%-----------%%%%%%%%%%-----------%%%%%%%%%%-----------%%%%%%%%%% Supplementary Material

\onecolumngrid
\begin{center}
\vspace{1cm}
\textbf{\large Deterministic generation of photonic entangled states using decoherence-free subspaces: Supplemental Material}\\[.2cm]

Oriol Rubies-Bigorda$^{1,2}$, Stuart J. Masson$^{3}$, Susanne F. Yelin$^{2}$, and Ana Asenjo-Garcia$^{3}$\\
[.1cm]
{\small \itshape
${}^1$Physics Department, Massachusetts Institute of Technology, Cambridge, Massachusetts 02139, USA\\
${}^2$Department of Physics, Harvard University, Cambridge, Massachusetts 02138, USA\\
${}^3$Department of Physics, Columbia University, New York, New York 10027, USA}

\newcommand{\beginsupplement}{%
        \setcounter{table}{0}
        \renewcommand{\thetable}{S\arabic{table}}%
        \setcounter{figure}{0}
        \renewcommand{\thefigure}{S\arabic{figure}}%
     }
\end{center}
\newcommand{\beginsupplement}{%
        \setcounter{table}{0}
        \renewcommand{\thetable}{S\arabic{table}}%
        \setcounter{figure}{0}
        \renewcommand{\thefigure}{S\arabic{figure}}%
     }
%\renewcommand{\thefigure}{S\arabic{figure}}
%%%%%%%%%%% Merge with supplemental materials %%%%%%%%%%
%%%%%%%%%% Prefix a "S" to all equations, figures, tables and reset the counter %%%%%%%%%%
\setcounter{equation}{0}
\setcounter{figure}{0}
\setcounter{table}{0}   
\setcounter{page}{1}
% \makeatletter
\renewcommand{\theequation}{S\arabic{equation}}
\renewcommand{\thefigure}{S\arabic{figure}}
\renewcommand{\bibnumfmt}[1]{[S#1]}
\renewcommand{\citenumfont}[1]{S#1}
\vspace{0.8 in}
\newcommand{\D}{\Delta}
\newcommand{\tD}{\tilde{\Delta}}
\newcommand{\K}{K_{PP}}
\newcommand{\bn}{\bar{n}_P}
\newcommand{\G}{\Gamma}
\newcommand{\LH}{\underset{L}{H}}
\newcommand{\HL}{\underset{H}{L}}
%\renewcommand{\thefigure}{S\arabic{figure}}
%%%%%%%%%% Prefix a "S" to all equations, figures, tables and reset the counter %%%%%%%%%%
\vspace{-1in}

\let\oldaddcontentsline\addcontentsline% Store \addcontentsline
\renewcommand{\addcontentsline}[3]{}% Make \addcontentsline a no-op
\section{Contents}
\vspace{-1cm}
%\printbibliography[heading=none]
\let\addcontentsline\oldaddcontentsline% Restore \addcontentsline

\tableofcontents

\section{I. Master equation for an ensemble of
quantum emitters coupled to a half-waveguide}
\label{SI: formalism}

We consider an ensemble of $N$ two-level emitters coupled to a half-waveguide. The Hamiltonian describing the full system is $\hat{H}_t = \hat{H}_S + \hat{H}_R + \hat{H}_\mathrm{int}$, with
\begin{align}
    \hat{H}_S &= \omega_0 \sum_n \hat{\sigma}_n^\dagger \hat{\sigma}_n, \\
    \hat{H}_R &= \int_0^\infty dk \omega_k \hat{b}_k^\dagger  \hat{b}_k , \\
    \hat{H}_\mathrm{int} &=  g \sum_n \int_0^\infty dk \sin(k x_n) (\hat{b}_k^\dagger +  \hat{b}_k) (\hat{\sigma}_n^\dagger + \hat{\sigma}_n),
\end{align}
where $\hat{\sigma}_n$ is the lowering operator of the two-level emitter at position $x_n$, $\hat{b}_k$ is the bosonic operator describing a waveguide mode with momentum $k = \omega/c$ and electric field profile $\sin (k x)$, and $g$ denotes the constant coupling strength between the waveguide and the emitters. The emitter-waveguide interaction in the interaction picture reads
\begin{align}
    \hat{\tilde{H}}_\mathrm{int} &= e^{i (\hat{H}_S + \hat{H}_R)} \hat{H}_\mathrm{int} e^{-i (\hat{H}_S + \hat{H}_R)} = g \sum_n \int_0^\infty dk \sin(k x_n) (\hat{b}_k^\dagger e^{i \omega_k t} +  \hat{b}_k e^{-i \omega_k t}) (\hat{\sigma}_n^\dagger e^{i \omega_0 t} + \hat{\sigma}_n e^{-i \omega_0 t}).
\end{align}

Formally integrating the von-Neumann equation of motion for the full system, $d\hat{\rho}/dt = - i [\hat{\tilde{H}}_\mathrm{int}, \hat{\rho}]$, allows for tracing over the electromagnetic degrees of freedom (reservoir R). We obtain the integro-differential equation describing the dynamics of the emitters (system S) \cite{CohenTanoudgi_book,Meystre_book}
\begin{equation}
    \frac{d}{dt} \hat{\rho}_S = - \int_0^t d\tau \mathrm{Tr}_R \left[\hat{\tilde{H}}_\mathrm{int} (t), \left[ \hat{\tilde{H}}_\mathrm{int} (\tau), \hat{\rho} (\tau) \right]  \right].
\end{equation}
We further apply the Born-Markov approximation, which assumes that the reservoir is (i) large compared to the system and therefore unchanged, and (ii) has a very short correlation time (\ie the reservoir has a very broad band). This is fulfilled if the time needed for a photon to propagate from one emitter to another, or to the mirror, is much smaller than the characteristic timescale of the emitters. Then, we can approximately factorize the density matrix, $\hat{\rho} (\tau) \approx \hat{\rho}_S (t) \otimes |0 \rangle \langle 0 |$, where the last term indicates that the waveguide field (reservoir) remains in the vacuum. Expanding the time integration to infinity according to the Markov approximation and using the identity $\int_0^\infty d \tau e^{i \omega \tau} = \pi \delta(\omega) + i \mathcal{P} \omega^{-1}$, where $\mathcal{P}$ denotes the Cauchy principal value, we obtain the final master equation for the emitters 
\begin{equation}
\label{eq: Master_equation_general}
    \frac{d\hat{\rho}}{dt} = -i \left[ \sum_{n,m} J_{nm}  \hat{\sigma}_n^\dagger \hat{\sigma}_m, \hat{\rho}\right] + \sum_{n,m} \frac{\Gamma_{nm}}{2} \left( 2 \hat{\sigma}_m \hat{\rho} \hat{\sigma}_n^\dagger - \hat{\sigma}_n^\dagger \hat{\sigma}_m \hat{\rho} - \hat{\rho} \hat{\sigma}_n^\dagger \hat{\sigma}_m \right).
\end{equation}
The coherent and dissipative interactions in Eq.~(\ref{eq: coh_dis_int}) result from the integrals 
\begin{align}
    \Gamma_{nm} &= \frac{\pi g^2}{c} \int_0^\infty d \omega \sin \left(\frac{\omega x_n}{c} \right) \sin \left(\frac{\omega x_m}{c} \right) \delta(\omega - \omega_0), \\
    J_{nm} &= - \frac{g^2}{c} \mathcal{P} \int_0^\infty d \omega \sin \left(\frac{\omega x_n}{c} \right) \sin \left(\frac{\omega x_m}{c} \right) \left( \frac{1}{\omega - \omega_0} + \frac{1}{\omega + \omega_0} \right).
\end{align}
The integral corresponding to the energy shift needs to be performed via contour integration by splitting the squared sinusoidal function into the components with positively and negatively oscillating exponents, respectively. Defining the decay rate $\gamma_0 = 2 \pi g^2/c$, we obtain the final form of the coherent and dissipative interactions
\begin{equation}
\label{eq: coh_dis_int}
    J_{nm} - i \frac{\Gamma_{nm}}{2} = - i\frac{ \gamma_0}{4} \left( e^{i k_0 |x_n - x_m|} - e^{i k_0 |x_n + x_m|} \right),
\end{equation}
where $k_0 = \omega_0 / c$ is the resonant wavenumber of light. Due to the presence of the mirror, the coherent and dissipative interactions between two emitters at positions $x_n$ and $x_m$ have two contributions. The first corresponds to a photon propagating directly from emitter $n$ to emitter $m$, thereby acquiring a phase proportional to their separation $|x_n - x_m|$. The second corresponds to a photon propagating from emitter $n$ to the mirror and back to emitter $m$ (or viceversa), thereby acquiring a phase proportional to the total distance traveled by the photon, $|x_n + x_m|$. Due to the mirror, an additional phase of $\pi$ is picked up by the photon upon reflection. Note also that the coherent and dissipative interactions for a waveguide would simply be given by the first contribution in Eq.~(\ref{eq: coh_dis_int}).

Importantly, the Lindbladian or dissipative part of the master equation can be written in terms of a single collective jump operator $\hat{S} =\sqrt{\gamma_0 /\Gamma_B} \sum_n \sin (k_0 x_n) \hat{\sigma}_n$ with decay rate $\Gamma_B = \gamma_0 \sum_n \sin^2 (k_0 x_n)$,
\begin{equation}
\label{eq: lindbladian_short}
    \mathcal{L}(\hat{\rho}) = \frac{\Gamma_{B}}{2} \left( 2 \hat{S} \hat{\rho} \hat{S}^{\dagger} - \hat{S}^{\dagger} \hat{S} \hat{\rho} - \hat{\rho} \hat{S}^{\dagger} \hat{S} \right).
\end{equation}
This jump operator represents the decay of excitations in the system through the emission of a photon towards the transmitting side of the waveguide.

Finally, we can also compute the total field in the half-waveguide as a function of time by deriving the equations of motion for the operator $\hat{b}_k$, $d \hat{b}_k /dt = i [\hat{H}_t, \hat{b}_k] = -i \omega_k \hat{b}_k - i g \sum_n \sin (k x_n) (\hat{\sigma}_n^\dagger + \hat{\sigma}_n)$. Formally integrating this equation and plugging it into the general expression for the electromagnetic field, $\hat{E}(x) = \int dk g \sin (k x) (\hat{b}_k + \hat{b}_k^\dagger)$, we obtain
\begin{equation}
\label{eq: SM_field_emitted}
    \hat{E}(x) = \Omega \sin (k_0 x) - i \frac{\gamma_0}{4} \sum_n \left( e^{i k_0 |x - x_n|} - e^{i k_0 |x + x_n|} \right) \hat{\sigma}_n,
\end{equation}
where $\Omega$ denotes a classical driving field in the half-waveguide. To the right of the emitters, the contribution from emitter $n$ simply reads $-2 i e^{i k_0 x} \sin (k_0 x_n)$. Splitting $\Omega \sin (k x)$ into the incident ($e^{-i k_0 x}$) and reflected components ($e^{i k_0 x}$), we can express the reflection coefficient as
\begin{equation}
    r = -1 + i \frac{\gamma_0}{\Omega} \sum_n \sin (k_0 x_n) \langle \hat{\sigma}_n \rangle.
\end{equation}
In the absence of emitters, the reflection coefficient corresponds to that of a mirror, \ie $r=-1$.

In this work, we consider the configuration where the emitters are located at $x_n = (n +1/4) \lambda_0$. The emitters then exhibits all-to-all dissipative interactions (\ie $J_{nm} = 0$ and $\Gamma_{nm} = \gamma_0$ $\forall n,m$), and the master equation~(\ref{eq: Master_equation_general}) simplifies to Eq.~(\textcolor{blue}{1}) of the main text.

\section{II. Matter gates in the decoherence-free subspace}
\label{SI: gates}

In this section, we present in detail the full set of quantum gates between the three dark states in the decoherence-free subspace (DFS), namely the ground state $\ket{G} \equiv |ggg\rangle$ and the single-excitation collective states $\ket{D} = (\ket{egg} - \ket{gge}) / \sqrt{2}$ and $\ket{A} = (\ket{egg} - 2\ket{geg} + \ket{gge}) / \sqrt{6}$. We consider the Hamiltonian in Eq.~(\textcolor{blue}{2}) of the main text with $\delta = -J$. In terms of the lowering and rising operators associated to the collective states, $\hat{\sigma}_B = (\hat{\sigma}_1 + \hat{\sigma}_2 + \hat{\sigma}_3)/\sqrt{3} \equiv \hat{S}$, $\hat{\sigma}_D = (\hat{\sigma}_1 - \hat{\sigma}_3)/\sqrt{2}$ and $\hat{\sigma}_A = (\hat{\sigma}_1 -2 \hat{\sigma}_2 + \hat{\sigma}_3)/\sqrt{6}$, it can be expressed as
\begin{equation}
    \hat{H}_{nH} = \left(\Delta + J - i \frac{3\gamma_0}{2}\right) \hat{\sigma}_B^\dagger \hat{\sigma}_B + \Delta \hat{\sigma}_D^\dagger \hat{\sigma}_D + (\Delta -2 J) \hat{\sigma}_A^\dagger \hat{\sigma}_A + \sum_{k \in \{ D,A,B \} } \left( \Omega_k e^{-i (\omega_L - \omega_0) t} \hat{\sigma}_k^\dagger + h.c. \right),
\end{equation}
where the non-Hermitian term $- i 3 \gamma_0 \hat{\sigma}_B^\dagger \hat{\sigma}_B /2$ describes the decay of atomic excitation via photon emission into the half-waveguide. In the absence of drive and as shown in Fig.~\textcolor{blue}{1}(b), the single-excitation eigenstates of the non-Hermitian Hamiltonian $\hat{H}_{nH}$ are $\ket{B} = \hat{\sigma}_B^\dagger \ket{G}$ (shifted from $\omega_0$ by $\Delta +J$ and exhibiting a decay rate $3 \gamma_0$), $\ket{D}$ (shifted by $\Delta$) and $\ket{A}$ (shifted by $\Delta - 2 J$). The two-excitation eigenstates are $\ket{S_D}$ (shifted from $2\omega_0$ by $2\Delta -J$ and exhibiting a decay rate $\gamma_0$), $\ket{\lambda_1} = \epsilon_1 \ket{S_B} + \epsilon_2 \ket{S_A}$ (shifted by $2\Delta -2J$ and exhibiting a decay rate $\approx \gamma_0$) and $\ket{\lambda_2} = \epsilon_2 \ket{S_B} - \epsilon_1 \ket{S_A}$ (shifted by $2\Delta +J$ and exhibiting a decay rate $\approx 4\gamma_0$). Here, we have defined the states $\ket{S_k} = \hat{S}^\dagger \ket{k} / || \hat{S}^\dagger \ket{k} ||$ for $k \in \{ D,A,B \}$. For the sake of brevity, we refrain from writting down the explicit form of $\epsilon_{1,2}$. Finally, the drive introduces couplings between states with different excitation numbers.

Arbitrary single-qutrit gates for a general state $\ket{\psi_{DFS}(t)} = d(t) e^{-i \omega_0 t} \ket{D} + g(t) \ket{G} + a(t) e^{-i \omega_0 t} \ket{A}$ in the DFS can be obtained by sequentially applying the following five operations, which enable arbitrary phase control and arbitrary rotations between two pairs of states~\cite{qutrit_1,qutrit_2,qutrit_3}.

\subsection{II.A. Arbitrary rotations between $\ket{D}$ and $\ket{G}$}

We drive the $\ket{G} \leftrightarrow \ket{D}$ transition on resonance, $\omega_0 = \omega_L$ and $\Delta=0$, by applying $\Omega_D = \Omega e^{i\phi}$ and $\Omega_B=\Omega_A=0$. For the individual emitters, this corresponds to the drive $\Omega_1 = - \Omega_3 = \Omega e^{i\phi}/\sqrt{2} $. This drive does not only couple $\ket{D}$ and $\ket{G}$ with strength $\Omega$, but also gives rise to the transitions $\ket{A} \leftrightarrow \ket{S_D}$ and $\ket{D} \leftrightarrow \ket{\lambda_{1,2}}$. Notably, the first process is off-resonant by $-J$, whereas the latter are off-resonant by $-2J$ and $J$. Additionally, the two-excitation bright states decay at rates ranging from $\gamma_0$ to $4 \gamma_0$. For weak drive compared to the combined effect of the off-resonance and decay rates, $\Omega^2 \ll \gamma_0^2 + J^2$, the bright states outside the DFS can be adiabatically eliminated. To leading order in the small parameter $\Omega^2 / (\gamma_0^2 + J^2)$, they project the state back into the DFS, resulting in the Hamiltonian
\begin{equation}
    \hat{H}_{DG} \approx -2 J \ket{A} \bra{A} + \Omega e^{i \phi} \ket{D} \bra{G} +  \Omega e^{-i \phi} \ket{G} \bra{D},
\end{equation}
where we have only included the terms involving the three dark states. After a time $T \sim \Omega^{-1} \gg 1/ \sqrt{J^2 + \gamma_0^2}$, these dynamics generate the arbitrary rotations $\mathcal{R}_{DG}$ given by Eq.~(\textcolor{blue}{3}) in the main text.

In reality, the coupling to the bright states induces small errors or corrections to $\mathcal{R}_{DG}$. To estimate this error, we note that the protocols for entangled photon generation only involve rotations between $\ket{D}$ and $\ket{G}$ when there is no amplitude in $\ket{A}$, that is, when the initial state reads $\ket{\psi_{DFS}(t=0)} = d_0 \ket{D} + g_0 \ket{G}$. The evolution of the state of the system including the two-excitation states $\ket{\lambda_{1,2}}$, i.e., $\ket{\psi(t)} = d(t) \ket{D} e^{-i\omega_0t} + g_0 \ket{G} + \lambda_1 e^{-2i\omega_0t} \ket{\lambda_1} + \lambda_2 e^{-2i\omega_0t} \ket{\lambda_2}$, reads
\begin{subequations}
    \begin{align}
        \dot{g}(t) & = - i \Omega e^{-i \phi} d(t), \\
        \dot{d}(t) & = - i \Omega e^{i \phi} g(t) - i \Omega e^{-i \phi} \xi_1^* \lambda_1(t) - i \Omega e^{-i \phi} \xi_2^* \lambda_2(t), \\
        \dot{\lambda}_1(t) & = -i (-2J - i \frac{\gamma_0}{2}) \lambda_1(t) - i \Omega e^{i \phi} \xi_1 d(t), \\
        \dot{\lambda}_2(t) & = -i (J - i \frac{4\gamma_0}{2}) \lambda_2(t) - i \Omega e^{i \phi} \xi_2 d(t),
    \end{align}
\end{subequations}
where the overlaps $\xi_{1,2} = \bra{\lambda_{1,2}} \sigma_d^\dagger \ket{D}$ take values $|\xi_{1,2}|^2 \in \{ 0.67,0.33 \}$ with $|\xi_1|^2 + |\xi_2|^2 = 1$ that depend on $J$ and $\gamma_0$. Note that we have assumed that the states $\ket{\lambda_{1,2}}$ do not decay to $\ket{B}$ and $\ket{A}$, but rather to states outside of the system. While this results in a lower bound for the achievable fidelities, it results in the correct scaling with the relevant system parameters, \ie $J$, $\gamma_0$ and the gate time $T$. For $\Omega \ll \sqrt{J^2 + \gamma_0^2}$, the rapidly evolving two-excitation bright states can be adiabatically eliminated by setting $d \lambda_{1,2}/dt =0$. Solving for the instantaneous values of $\lambda_{1,2}(t)$ as a function of $d(t)$, we finally obtain the effective equation for the amplitude in the dark state $\ket{D}$,
\begin{equation}
    \dot{d}(t) = -i \left( \delta_d - i \frac{\gamma_d}{2} \right) d(t) - i \Omega e^{i \phi} g(t),
\end{equation}
where the energy shift $\delta_d$ and decay rate $\gamma_d$ induced by the coupling to the two-excitation states are
\begin{equation}
    \gamma_d = \Omega^2 \gamma_0 \left( \frac{|\xi_1|^2}{4J^2 + \gamma_0^2/4} + \frac{4|\xi_2|^2}{J^2 +4 \gamma_0^2}\right), \quad \quad \quad \delta_d = \Omega^2 J \left(  \frac{2|\xi_1|^2}{4J^2 + \gamma_0^2/4} - \frac{|\xi_2|^2}{J^2 +4 \gamma_0^2} \right).
\end{equation}
Applying a global detuning $\Delta = - \delta_d$ such that the drive is still on resonance with the $\ket{G} \leftrightarrow \ket{D}$ transition and taking $\phi = -\pi/2$, an initial state $\ket{\psi(0)} = d_0 \ket{D} + g_0 \ket{G}$ performs a rotation around the $y$ axis of the qubit defined by $\ket{D}$ and $\ket{G}$ 
\begin{equation}
    \begin{pmatrix}
    d(t)\\
    g(t)
    \end{pmatrix} = \tilde{U}(t) \begin{pmatrix}
    d_0\\
    g_0
    \end{pmatrix} =  e^{-\gamma_d t /4}  
    \begin{pmatrix}
      \cos (\Omega_\mathrm{eff}t) - \frac{\gamma_d}{4\Omega_\mathrm{eff}} \sin (\Omega_\mathrm{eff}t)  & -\frac{\Omega}{\Omega_\mathrm{eff}} \sin (\Omega_\mathrm{eff}t) \\
     \frac{\Omega}{\Omega_\mathrm{eff}} \sin (\Omega_\mathrm{eff}t) &   \cos (\Omega_\mathrm{eff}t) + \frac{\gamma_d}{4\Omega_\mathrm{eff}} \sin (\Omega_\mathrm{eff}t) 
    \end{pmatrix}
    \begin{pmatrix}
    d_0\\
    g_0
    \end{pmatrix} ,
\end{equation}
where we have defined the effective Rabi frequency $\Omega_\mathrm{eff} = \sqrt{\Omega^2 - \gamma_d^2/16}$. For $\Omega_\mathrm{eff} T = \pi/4$, the evolution $\ket{\psi(T)} = d(t) e^{-i \omega_0 T} \ket{D} + g(t) \ket{G}$ approximately results in a $\pi/2$ rotation around the $y$ axis, characterized by the unitary
\begin{equation}
    \tilde{U}(T) =  \frac{e^{-\gamma_d T /4}  }{2}
    \begin{pmatrix}
      1 - \frac{\gamma_d}{4\Omega_\mathrm{eff}}   & -\frac{\Omega}{\Omega_\mathrm{eff}} \\
     \frac{\Omega}{\Omega_\mathrm{eff}}  &   1 + \frac{\gamma_d}{4\Omega_\mathrm{eff}} 
    \end{pmatrix} ,
\end{equation}
The ideal unitary $U$, achieved in the case where $\gamma_d = 0$ and thus $\Omega_\mathrm{eff} = \Omega$, reads 
\begin{equation}
    U(T) =  \frac{e^{-\gamma_d T /4}  }{2}
    \begin{pmatrix}
      1    & -1 \\
     1  &   1 
    \end{pmatrix} .
\end{equation}
The average gate fidelity can be computed as~\cite{fidelity_3,fidelity_1,fidelity_2}
\begin{equation}
    \mathcal{F} = \frac{ 1 + d^{-1} |\mathrm{tr} ( U^\dagger \tilde{U})|^2}{ d+1} , 
\end{equation}
where $d=2$ describes the dimension of the Hilbert space.
To lowest order in the small parameter $\gamma_d/\Omega$, the fidelity is found to be
\begin{equation}
    \mathcal{F} = 1 - \frac{\pi \gamma_d}{12\Omega} + \mathcal{O} \left(\frac{\gamma_d^2}{ \Omega^2}\right) . 
\end{equation}
Noting that $T \approx \pi/4\Omega$ and $\gamma_d \propto \Omega^2 \gamma_0 /(J^2 + \gamma_0^2)$, one readily finds that the error of the quantum gate scales as
\begin{equation}
    \epsilon = 1 - \mathcal{F} \sim \mathcal{C} \frac{1}{T} \frac{\gamma_0}{J^2 + \gamma_0^2},
\end{equation}
where $\mathcal{C}$ is a constant. We thus find that the error for $J\gg \gamma_0$ scales as $\epsilon \propto \gamma_0 / T J^{2}$.

\subsection{II.B. Arbitrary rotations between $\ket{G}$ and $\ket{A}$}

We drive the $\ket{G} \leftrightarrow \ket{A}$ transition on resonance, $\omega_0 = \omega_L$ and $\Delta=2J$, by applying $\Omega_A = \Omega e^{i\phi}$ and $\Omega_B=\Omega_D=0$. For the individual emitters, this corresponds to the drive $\Omega_1 = - \Omega_2/2 = \Omega_3 = \Omega e^{i\phi}/\sqrt{6} $. Apart from coupling $\ket{G}$ and $\ket{A}$ with strength $\Omega$, this drive also gives rise to the transitions $\ket{D} \leftrightarrow \ket{S_D}$ and $\ket{A} \leftrightarrow \ket{\lambda_{1,2}}$. The first process is off-resonant by $-J$, while the latter are off-resonant by $-2J$ and $-5J$. Again, the off-resonant drive and the decay of the bright states projects the evolution of the system into the DFS for small drives $\Omega^2 \ll \gamma_0^2 + J^2$. To leading order, the relevant terms in the Hamiltonian read
\begin{equation}
\label{eq:SI_Ham_rot_ga}
    \hat{H}_{GA} \approx 2 J \ket{D} \bra{D} + \Omega e^{i \phi} \ket{G} \bra{A} +  \Omega e^{-i \phi} \ket{A} \bra{G},
\end{equation}
After a time $T \sim \Omega^{-1} \gg 1/ \sqrt{J^2 + \gamma_0^2}$, $\hat{H}_{GA}$ results in the gate 
\begin{equation}
\label{eq:rotation_GA_SM}
    \mathcal{R}_{GA}(\theta,\phi,\chi) = \begin{pmatrix}
 e^{ -i \chi} & 0 & 0 \\
0 & \cos(\theta) & - i e^{i \phi} \sin (\theta) \\
0 & - i e^{-i \phi} \sin (\theta) & \cos(\theta)
\end{pmatrix},
\end{equation}
which implements arbitrary rotations between $\ket{G}$ and $\ket{A}$ while $\ket{D}$ acquires a phase $\chi = -2JT$. The protocols for entangled state generation rely on applying $\mathcal{R}_{GA}$ on the initial state $\ket{\psi(0)} = d_0 \ket{D} + g_0 \ket{G}$. As a result, the gates act on the whole DFS of dimension $d=3$. Following a similar derivation as for the rotations between $\ket{D}$ and $\ket{G}$, one finds the same scaling of the gate infidelity to leading order, $\epsilon \propto \gamma_0 / [T (J^2 + \gamma_0^2)]$.   

\subsection{II.C. Phase control of $\ket{A}$}

Applying only the coherent coupling $J$ results in the Hamiltonian $\hat{H}_A = -2 J \ket{A} \bra{A}$, where we have only showed the terms involving the three dark states. Applying $\hat{H}_A$ for a time $T$ results in the phase gate 
\begin{equation}
    \mathcal{P}_A (\phi) = \exp(- i \hat{H}_A T) = \begin{pmatrix}
1 & 0 & 0\\
0 & 1 & 0 \\
0 & 0 & e^{i\phi}
\end{pmatrix}
\end{equation}
with $\phi = 2 J T$. For large $J$, $\mathcal{P}_A(\phi)$ can be applied in a time $T \ll \gamma_0^{-1}$.  

\subsection{II.D. Phase control of $\ket{D}$}

Applying only an equal detuning $\Delta$ to all emitters gives rise to the Hamiltonian $\hat{H}_D = \Delta (\ket{D}\bra{D}+\ket{A}\bra{A})$ within the DFS. The resulting phase gate is
\begin{equation}
    \mathcal{P}_D (\phi) = \exp(- i \hat{H}_d T) = \begin{pmatrix}
e^{-i\phi} & 0 & 0\\
0 & 1 & 0 \\
0 & 0 & e^{-i\phi}
\end{pmatrix}
\end{equation}
with $\phi = \Delta T$. Together with $\mathcal{P}_A$, it allows to control the phase of state $\ket{D}$. Again, $\mathcal{P}_D(\phi)$ can be applied in a time $T \ll \gamma_0^{-1}$ by considering $\Delta \gg \gamma_0$. 

\subsection{II.E. Phase control of $\ket{G}$}

Control over the phase of the ground state is attained by shifting the resonance frequencies of the dark states via a far off-resonant drive with Rabi frequency $\Omega_n = \Omega_B / \sqrt{3}$  equal for all emitters $n \in \{ 1,2,3 \}$. For $\Delta = J = \Omega_D = \Omega_A = 0$ and defining $\delta \omega = \omega_L - \omega_0$, the Hamiltonian (in the rotating frame of the drive) reads
\begin{equation}
    \hat{H} = \left( -\delta \omega - i \frac{3 \gamma_0}{2} \right) \ket{B} \bra{B} + \left( -\delta \omega - i \frac{ \gamma_0}{2} \right) \left( \ket{D} \bra{D} + \ket{A} \bra{A} \right) + \Omega_B \left( \ket{G}\bra{B} + \frac{1}{\sqrt{3}}\ket{D}\bra{S_D} + \frac{1}{\sqrt{3}} \ket{A}\bra{S_A} + h.c. \right). 
\end{equation}
The resulting equation of motion for the amplitudes in $\ket{G}$ and $\ket{B}$, for example, are $\dot{g} = - i \Omega_B b$ and $\dot{b}=- i \Omega_B - i (-\delta \omega - i 3\gamma_0/2)b$. For $|\delta \omega| \gg \Omega_D,\gamma_0$, the bright state can be adiabatically eliminated. To leading order, this results in a phase shift of the ground state, $\dot{g} \approx - i \Omega_B^2 g / \delta\omega$. A similar treatment results in the equations $\dot{c} \approx - i \Omega_B^2 c / 3\delta\omega$ for $c \in \{d,a \}$. After a time $T$, the corresponding phase gate thus reads
\begin{equation}
    \mathcal{P}_{G}(\phi) = \begin{pmatrix}
e^{-i\phi/3} & 0 & 0\\
0 & e^{-i\phi} & 0 \\
0 & 0 & e^{-i\phi/3}
\end{pmatrix}
\end{equation}
with $\phi = \Omega^2 T / \delta \omega $. Together with $\mathcal{P}_A$ and $\mathcal{P}_D$, $\mathcal{P}_G$ allows to control the phase of the ground state $\ket{G}$.

\section{III. Single photon emission}
\label{SI: emission}

In this section, we present the emission of single photons with arbitrary temporal wavepackets $\psi_\mathrm{ph}(t)$ (such that $\int |\psi_\mathrm{ph}(\tau)|^2 d \tau  =1$) from the single-excitation collective dark state $\ket{A} = (\ket{egg} - 2\ket{geg} + \ket{gge}) / \sqrt{6}$. This is achieved by coupling $\ket{A}$ to the single-excitation bright state $\ket{B} = (\ket{egg} + \ket{geg} + \ket{gge}) / \sqrt{3}$ via a detuning $\delta \neq -J$ applied on the second emitter. In particular, in the absence of drives (\ie $\Omega_k = 0$) and setting $\Delta = - 4 J$ and $\delta = 8 J$, the Hamiltonian for the states in the single-excitation manifold reads
\begin{equation}
    \hat{H}_\mathrm{ph}(t) = - 4 J(t) \ket{D} \bra{D} - i \frac{3\gamma_0}{2} \ket{B} \bra{B} - 3 \sqrt{2} J(t) \left( \ket{A} \bra{B} + \ket{B} \bra{A} \right),
\end{equation}
where $J(t)$ may vary over time. Applying Schr\"{o}dinger's equation, we readily find that the evolution of a general state $\ket{\psi (t)} = d(t) e^{- i \omega_0 t} \ket{D} + a(t) e^{- i \omega_0 t} \ket{A} + b(t) e^{- i \omega_0 t} \ket{B}$ is given by
\begin{subequations}
\label{eq: SI_emission_amplitudes}
    \begin{align}
        \dot{d}(t) & = i 4 J(t) d(t), \\
        \dot{a}(t) & = i 3 \sqrt{2} J(t) b(t), \\
        \dot{b}(t) & = - \frac{3 \gamma_0}{2} b(t) + i 3 \sqrt{2} J(t) a(t).
    \end{align}
\end{subequations}
The wavepacket of the emitted photon, \ie the amplitude of the emitted electromagnetic field over time, only depends on the time-dependent amplitude $b(t)$ of the bright state $\ket{B}$, and can be obtained from Eq.~(\ref{eq: SM_field_emitted}) by setting the strength of the classical driving field to zero ($\Omega=0$)~\cite{Photon_wavepacket,Photon_wavepacket2}
\begin{equation}
\label{eq: SI_photon_amplitude}
    \psi_\mathrm{ph}(t) = \sqrt{3 \gamma_0} b(t).
\end{equation}
Note that this equation implies that the fraction of the photon emitted per unit of time, $|\psi_\mathrm{ph}(t)|^2$, is equal to the population of the bright state multiplied by its decay rate, $3 \gamma_0 |b(t)|^2$, as expected. Before proceeding, we assume the initial amplitude in state $\ket{A}$ to be imaginary. For a real coherent coupling $J$, Eq.~(\ref{eq: SI_emission_amplitudes}) enforces that $a(t)$ remains imaginary and $b(t)$ real during the whole evolution, thereby giving rise to a photon with a wavepacket $\psi_\mathrm{ph}(t)$ contained in the reals. 

Crucially, for an initial state $\ket{\psi(0)} = d_0 \ket{D} - i |a_0| \ket{A}$ with $|d_0|^2 + |a_0|^2=1$, the emission of a photon only occurs if the system is initially in $\ket{A}$. As a result, time evolution under Eq.~(\ref{eq: SI_emission_amplitudes}) gives rise to conditional photon emission and leads to a final state
\begin{equation}
    \ket{\psi(T_\mathrm{em})} = d_0 e^{i 4 \int_{0}^{T_\mathrm{em}} J(\tau) d\tau} e^{- i \omega_0 T_\mathrm{em}} \ket{D} \otimes \ket{0_\mathrm{ph}} + |a_0| \ket{G} \otimes \ket{1_\mathrm{ph}}, 
\end{equation}
where $T_\mathrm{em}$ denotes the total duration of the emission process (which is limited by the decay rate of $\ket{B}$ into the waveguide, $T_\mathrm{em} \gtrsim \gamma_0^{-1}$), and $\ket{0_\mathrm{ph}}$ and $\ket{1_\mathrm{ph}}$ denote the states with zero or one photon with wavepacket $\psi_\mathrm{ph}(t)$.  

In what follows, we discuss how to engineer different temporal profiles $\psi_\mathrm{ph}(t)$ by controlling $J(t)$. Since $\ket{D}$ is decoupled from the emission process, we consider for simplicity the initial state $\ket{\psi(0)} = - i \ket{A}$.

\begin{itemize}
    \item[(i)] For constant coherent coupling $J^2 > \gamma_0^2/32$, the photon wavepacket is an exponentially decaying sinusoidal function
\begin{equation}
    \psi_\mathrm{ph}(t) = \frac{J\sqrt{3 \gamma_0}}{\sqrt{J^2-\gamma_0^2/32}} \sin \left( t \sqrt{18J^2 - 9 \gamma_0^2 /16 } \right) e^{- 3 \gamma_0 t/4} \Theta(t),
\end{equation}
whereas for constant $J^2 < \gamma_0^2/32$, the oscillations are fully damped 
\begin{equation}
\label{eq: wavepacket_ctsmallJ}
    \psi_\mathrm{ph}(t) = \frac{J\sqrt{3 \gamma_0}}{\sqrt{\gamma_0^2/32  - J^2} } \sinh \left( t \sqrt{ 9 \gamma_0^2 /16 -18J^2 } \right) e^{- 3 \gamma_0 t/4} \Theta(t).
\end{equation} 
Here, $\Theta(t)$ represents the heaviside function, \ie $\Theta(t) = 1$ for $t\geq 0$ and $\Theta(t) = 0$ for $t < 0$. \\

\item[(ii)] Generating an arbitrary photon wavepacket $\psi_\mathrm{ph}(t)$ requires to optimize the coherent coupling over time. This can be easily achieved by discretizing the emission process in small time steps of duration $\delta t$~\cite{Subradiance_storage_control}, such that the atomic and photonic amplitudes in Eqs.~(\ref{eq: SI_emission_amplitudes}) and~(\ref{eq: SI_photon_amplitude}) in the $k$-th step read
\begin{subequations}
    \begin{align}
        \psi_\mathrm{ph}^{(k)} & = \sqrt{3 \gamma_0} b^{(k)}, \\
        b^{(k)} & = b^{(k-1)} + \delta t \left( - \frac{3 \gamma_0}{2} b^{(k-1)} - 3 \sqrt{2} J^{(k-1)} \mathrm{Im} \{ a^{(k-1)} \} \right), \\
        \mathrm{Im} \{ a^{(k)} \} & = \mathrm{Im} \{ a^{(k-1)} \} + 3 \sqrt{2} \delta t J^{(k-1)} b^{(k-1)},
    \end{align}
\end{subequations}
where we have used the fact that $a(t)$ is imaginary for all times if the photon wavepacket has real amplitudes. From these equations, we can readily obtain the coherent coupling at step $k-1$, $J^{(k-1)}$, that results in the desired photon wavepacket at step $k$, $\psi_\mathrm{ph}^{(k)}$,
\begin{equation}
\label{eq: optimal_coupling}
    J^{(k-1)} = \frac{1}{3 \sqrt{2} \delta t \mathrm{Im} \{ a^{(k-1)} \} } \left(  b^{(k-1)} \left( 1 - \frac{3 \gamma_0}{2} \delta t  \right) - \frac{\psi_\mathrm{ph}^{(k)}}{ \sqrt{3 \gamma_0}}  \right).
\end{equation}
This simple protocol allows to find the control sequence $J(t)$ that generates arbitrary photon wavepackets with large fidelity, provided that they are continuous, have zero amplitude at $t=0$ and that their duration $T_{em}$ is longer than the inverse decay rate $\sim \gamma_0^{-1}$. %We refer the reader to Ref.~\cite{ours_PRA_waveguide} for a detailed discussion on specific examples and the errors associated to the photon emission process. 

A simple analytic form of $J(t)$ can be derived in the limit where the emission rate is much larger than the coupling between $\ket{A}$ and $\ket{B}$, $J(t) \ll \gamma_0 $, for a Gaussian wavepacket
\begin{equation}
\label{eq: wavepacket_gaussian}
    \psi_\mathrm{ph} (t) = \frac{1}{\tau^{1/2} \pi^{1/4} } e^{- (t - t_0)^2/2 \tau^2}.
\end{equation}
In that case, the bright state can be adiabatically eliminated by setting $d b(t) / dt =0$. Formally integrating the resulting equation for the derivative of the amplitude in $\ket{A}$, one readily finds
\begin{equation}
\label{eq: SI_gaussian_adiabatic}
    \psi_\mathrm{ph} (t) = \sqrt{\gamma_\mathrm{eff} } e^{- \int_0^{t} dt' \gamma_\mathrm{eff} (t') /2}, 
\end{equation}
where we have defined the effective decay rate $\gamma_\mathrm{eff}(t) = 24 J(t)^2 / \gamma_0$. Comparing  Eq.~(\ref{eq: wavepacket_gaussian}) and Eq.~(\ref{eq: SI_gaussian_adiabatic}), one can finally obtain an analytical expression for the effective decay rate~\cite{Pichler_cluster}
\begin{equation}
    \gamma_\mathrm{eff}(t) = \frac{2 e^{- (t - t_0)^2 / \tau^2}}{\tau \sqrt{\pi} \left(1  - \mathrm{erf} ((t - t_0) / \tau) \right)},
\end{equation}
where $\mathrm{erf}(x) = 2 \int_0^x ds\, e^{-s^2} / \sqrt{\pi}$ denotes the Gaussian error function.

\end{itemize}

\section{IV. Light-matter gates}
\label{SI: light_gates}

The generation of entangled states of light requires entangling gates between the matter qubit $\ket{\psi_M(t)} = d(t) e^{-i \omega_t} \ket{D} + g(t) \ket{G}$ and the photonic qubits (\ie the absence or presence of a photon in a certain time bin). Here, we describe in detail how to attain the controlled photon emission (CPE) gate required for GHZ and cluster state generation, as well as the additional CZ gate needed to produce two-dimensional cluster states.

\subsection{IV.A. Conditional photon emission (CPE) gate}

The CPE gate for entangled state generation is always applied between the matter qubit and the $k$-th photonic qubit in the ground state, \ie in the state $\ket{0_k}$ without a photon. It corresponds to a conditional emission gate from the matter qubit,
\begin{equation}
\label{eq: CPE_gate}
    d_0 \ket{D} \otimes |0_k\rangle + g_0 \ket{G} \otimes |0_k\rangle \rightarrow d_0 e^{-i \omega_0 T_p} \ket{D} \otimes |0_k\rangle + g_0 \ket{G} \otimes |1_k\rangle, 
\end{equation}
where $\ket{1_k}$ denotes the state with one photon in the $k$-th time bin and $T_p$ is the gate duration. This operation can be engineered in three simple steps. First, the amplitude in the ground state is transferred to the single-excitation dark state $\ket{A}$ by applying $\mathcal{R}_{GA}(\pi/2,0,2 J_{GA} T_{GA})$, which results in the state $d_0 e^{- 2 i J_{GA} T_{GA} } e^{-i \omega_0 T_{GA}} \ket{D} \otimes |0_k\rangle - i g_0 e^{-i \omega_0 T_{GA}} \ket{A} \otimes |0_k\rangle$ after a gate time $T_{GA}$ and under a constant coherent exchange interaction $J_{GA}$. Second, a detuning is applied to the second emitter only, which couples $\ket{B}$ to $\ket{A}$ (but not to $\ket{D}$) and thereby generates a conditional photon of duration $T_{em}$. The resulting state reads, $d_0 e^{i \xi} e^{-i \omega_0 (T_{GA} + T_{em})} \ket{D} \otimes \ket{0_k} + g_0 \ket{G} \otimes \ket{1_k}$. Here, we have defined the phase $\xi = - 2 J_{GA} T_{GA} + 4 \int_0^{T_{em}} J(t') dt'$, where $J(t)$ is the sequence applied during the emission process to obtain a target photon wavepacket $\psi_{ph}(t)$. Finally, performing the phase gate $\mathcal{P}_D(\xi)$ by applying a constant detuning $\Delta$ over a time $T_D = \xi / |\Delta|$ corrects the additional phase acquired by $\ket{D}$ and completes the CPE gate in Eq.~(\ref{eq: CPE_gate}) with $T_p = T_{GA} + T_{em} + T_D$.  \\

\subsection{IV.B. CZ gate}
  
To apply a controlled phase gate on the photonic qubit emitted at step $k$, the photonic qubit is reflected by a switchable mirror placed at the transmitting end of the half-waveguide~\cite{Painter_states}. The subsequent scattering with the collective system allows to engineer a CZ gate that only performs a sign flip to the state $\ket{G} \otimes \ket{1_k}$. We assume that the bandwidth (or inverse duration) of the photon is much smaller than the decay rate of the collective bright state, such that the emitters are weakly driven. In this regime, the phase $r$ picked by the incoming photon upon reflection from the switchable mirror and the collective system is
\begin{equation}
    r = 1 - i \frac{3 \gamma_0}{\Omega_B} \langle \hat{\sigma}_B \rangle,
\end{equation}
where $\Omega_B$ is the small Rabi frequency associated to the incoming photon, and $\hat{\sigma}_B = \hat{S} = (\hat{\sigma}_1 + \hat{\sigma}_2+ \hat{\sigma}_3)/\sqrt{3}$ corresponds to the jump operator of the collective system into the waveguide.  

Let us first study the case where the matter qubit is initially in the ground state $\ket{G}$ and the incoming photon weakly excites the single-excitation bright state $\ket{B}$. For $\delta = -J$ (such that the single-excitation dark states remain eigenstates of the Hamiltonian interaction) and for weak drive with frequency $\omega_L = \omega_0 $ and strength $\Omega_B \ll | \Delta + J - i 3 \gamma_0/2|$, the amplitudes in the states $\ket{G}$ and $\ket{B}$ are
\begin{subequations}
    \begin{align}
        \dot{g}(t) &= - i  \Omega_B b(t), \\
        \dot{b}(t) &= - i (\Delta + J - i 
        \frac{3\gamma_0}{2}) - i  \Omega_B g(t).
    \end{align}
\end{subequations}
Due to the weak drive, the single-excitation bright state can be adiabatically eliminated. To leading order in the small parameter $|\Omega_B / (\Delta + J - i 3 \gamma_0/2)|$, the amplitude in the ground state remains $g(t) \approx 1$, while the amplitude in $\ket{B}$ reads $b(t) \approx -  \Omega_B / ( \Delta +J - i 3 \gamma_0/2)$. Noting that $\langle \hat{\sigma}_B \rangle = b$, we readily find the reflection coefficient for the ground state
\begin{equation}
\label{eq: reflection_coeff_g}
    r_G = 1 - 2 \frac{3 \gamma_0}{3 \gamma_0 + 2 i (J + \Delta)}.
\end{equation}
Similarly, the photon couples the single-excitation dark state $\ket{D}$ to the two-excitation bright state $\ket{S_D}$. Following an analogous derivation for the matter qubit initially in $\ket{D}$, we find the reflection coefficient
\begin{equation}
\label{eq: reflection_coeff_d}
    r_D = 1 - 2 \frac{\gamma_0}{ \gamma_0 + 2 i ( \Delta - J)}.
\end{equation}
As expected, the reflection coefficients $r_G$ and $r_D$ are respectively dictated by the decay rates of the collective states $\ket{B}$ and $\ket{S_D}$ and the off-resonance of the corresponding transitions. Note that the effect of the two mirrors (\ie the mirror forming the half-waveguide and the switchable mirror to reflect the photonic qubits back to the collective system) cancels out, such that a resonant photon acquires a phase flip ($r = - 1$) whereas a far off-resonant one does not ($r=1$).

The CZ gate is obtained by driving on resonance with the $\ket{G} \leftrightarrow \ket{B}$ transition, \ie for $\Delta = -J$. In that case, $r_g =-1$ and the photon acquires a phase flip when scattering of $\ket{G}$. When the matter qubit is in $\ket{D}$, however, the reflection coefficient $r_d = -e^{i \phi}$ is given by the phase $\phi = 2 \arctan (4 J / \gamma_0) $, which approaches $r_d \rightarrow 1$ for $J \gg \gamma_0$. The resulting ideal gate for the $k$-th photon reads
\begin{equation}
    \mathrm{CZ} (J) = \ket{G} \bra{G} \otimes \left( \ket{0_k} \bra{0_k} -  \ket{1_k} \bra{1_k} \right) + \ket{D} \bra{D} \otimes \left( \ket{0_k} \bra{0_k} + e^{2i \arctan (4J/\gamma_0)}  \ket{1_k} \bra{1_k} \right),
\end{equation}
which implements the CZ gate in the limit $J \gg \gamma_0$.

Due to the finite bandwidth of the photons (\ie their frequency distribution around $\Delta = - J$), the scattered wavepacket from $\ket{G}$, $\tilde{\psi}_\mathrm{ph}^{(G)}(t)$, and $\ket{D}$, $\tilde{\psi}_\mathrm{ph}^{(D)}(t)$, are not identical to the incoming wavepacket, $\psi_\mathrm{ph}(t)$. By means of the Fourier decomposition of the incoming wavepacket,
\begin{equation}
    \Psi_\mathrm{ph} (\omega) = \frac{1}{\sqrt{2\pi}} \int_{-\infty}^\infty\, d \omega \,\psi_\mathrm{ph} (t) \, e^{i \omega t},
\end{equation}
we readily compute the scattered wavepacket as
\begin{equation}
    \tilde{\psi}_\mathrm{ph}^{(\lambda)} (t) = \frac{1}{\sqrt{2\pi}} \int_{-\infty}^\infty d \omega\, \Psi_\mathrm{ph} (\omega)\, r_\lambda (\omega) \,e^{-i \omega t},
\end{equation}
where $\lambda \in \{ G, D \} $ represents the scattering process from each collective state. The frequency-dependent reflection coefficients for $\ket{G}$ and $\ket{D}$ are obtained from Eq.~(\ref{eq: reflection_coeff_g}) and (\ref{eq: reflection_coeff_d}) by substituting $\Delta \rightarrow \omega - J$. That is, given an incoming wavepacket $\psi_\mathrm{ph}(t)$, the collective system performs the quantum operation
\begin{equation}
    \tilde{U} = \ket{G} \bra{G} \otimes \left( \ket{0_k} \bra{0_k} +  \ket{1_k, \tilde{\psi}_\mathrm{ph}^{(G)}} \bra{1_k, \psi_\mathrm{ph} } \right) + \ket{D} \bra{D} \otimes \left( \ket{0_k} \bra{0_k} +  \ket{1_k, \tilde{\psi}_\mathrm{ph}^{(D)}} \bra{1_k, \psi_\mathrm{ph} } \right).
\end{equation}
The average gate fidelity can be computed as~\cite{fidelity_3,fidelity_1,fidelity_2}
\begin{equation}
    \mathcal{F} = \frac{ 1 + d^{-1} |\mathrm{tr} ( U^\dagger \tilde{U})|^2}{ d+1} , 
\end{equation}
where $d=4$ describes the dimension of the Hilbert space and $U$ implements the perfect CZ gate, 
\begin{equation}
\label{eq: CZ_gate_ideal}
    U = \ket{G} \bra{G} \otimes \left( \ket{0_k} \bra{0_k} -  \ket{1_k, \psi_\mathrm{ph} } \bra{1_k, \psi_\mathrm{ph} } \right) + \ket{D} \bra{D} \otimes \left( \ket{0_k} \bra{0_k} +  \ket{1_k, \psi_\mathrm{ph} } \bra{1_k, \psi_\mathrm{ph} } \right).
\end{equation}

Defining the overlap between the incoming and scattered wavepackets as
\begin{equation}
    O_\lambda = \int_{-\infty}^\infty d t \,\psi_\mathrm{ph}^* (t) \,\tilde{\psi}_\mathrm{ph}^{(\lambda)} (t),
\end{equation} 
the average fidelity reads
\begin{equation}
    \mathcal{F} = \frac{1}{5} + \frac{1}{20} | 2 - O_G + O_D|^2.
\end{equation}

(a) For the Gaussian wavepacket given by Eq.~(\ref{eq: wavepacket_gaussian}), obtained by dynamically controlling the coupling between the auxiliary dark state $\ket{A}$ and the bright state $\ket{B}$ during the emission process, the overlaps are
\begin{subequations}
\begin{align}
    O_G & = 1 - \sqrt{\pi} 3 \gamma_0 \tau e^{(3 \gamma_0 \tau /2)^2} \left( 1 - \mathrm{erf} \left( \frac{3 \gamma_0 \tau}{2} \right)\right) ,\\
    O_D & = 1 - \sqrt{\pi} \gamma_0 \tau e^{ \tau^2 (\gamma_0 - 4 i J )^2 /4} \left( 1 - \mathrm{erf} \left( \tau \frac{ \gamma_0 - 4 i J}{2} \right)\right).
\end{align} 
\end{subequations}

Large overlaps are obtained when the duration $\sim \tau$ of the wavepacket is much larger than the lifetime of the bright states $\sim \gamma_0^{-1}$, such that the photon effectively acts as a weak drive. This corresponds to the limit where the bandwidth of the photon, $\mathcal{B} = \tau^{-1}$, is much narrower than the bright state decay rates, that is, $\mathcal{B}/\gamma_0 \ll 1$. Then, the average gate fidelity reads 

\begin{equation}
    \mathcal{F} = 1 - \frac{4}{5} \frac{\gamma_0^2}{\gamma_0^2 + 16 J^2} - \frac{8}{45} \frac{\mathcal{B}^2}{\gamma_0^2} + \mathcal{O} \left(\frac{\mathcal{B}^4}{ \gamma_0^4}\right) + \mathcal{O} \left(\frac{\mathcal{B}^2}{ \gamma_0^2 } \frac{\gamma_0^3}{J^3}\right).
\end{equation}

The first error term arises from the phase $\phi = 2 \arctan(4J/\gamma_0) \neq \pi$ at finite $J/\gamma_0$. The second error term arises from the finite bandwidth of the photon scattering from $\ket{G}$. Note that the error associated to the finite bandwidth of the photon scattering from $\ket{D}$ is suppressed by an additional factor $\gamma_0^3/J^3$.

(b) For the photon wavepacket given by Eq.~(\ref{eq: wavepacket_ctsmallJ}) under the substitution $J \rightarrow \tilde{J}$, obtained by applying a constant coupling $\tilde{J} < \gamma_0 / \sqrt{32}$ during the emission process, the overlaps are found to be 
\begin{subequations}
    \begin{align}
        O_G & = - \frac{ \gamma_0^2 - 4 \tilde{J}^2}{\gamma_0^2 + 4 \tilde{J}^2},\\
        O_D & = - \frac{\left( \gamma_0 + 4 i J \right) \left( \gamma_0 - i J \right) - 18 \tilde{J}^2}{\left( \gamma_0 - 4 i J \right) \left( \gamma_0 - i J \right) + 18 \tilde{J}^2}.
    \end{align} 
\end{subequations}

In this case, the photon wavepacket is asymmetric. It is centered around $t_\mathrm{av} = \int_{-\infty}^\infty |\psi_\mathrm{ph}(t)|^2 t dt = 2/(3 \gamma_0) + \gamma_0/(24\tilde{J}^2)$ and has a temporal width $\tau^2 = \int_{-\infty}^\infty |\psi_\mathrm{ph}(t)|^2 (t-t_\mathrm{av})^2 dt \approx 5 \gamma_0^2 / (576 \tilde{J}^4)$. The overlap is maximal for large temporal widths $\tau$, and therefore for small values of $\tilde{J} \ll \gamma_0$. In this limit, the fidelity of the CPE gate reads
\begin{equation}
    \mathcal{F} = 1 - \frac{4}{5} \frac{\gamma_0^2}{\gamma_0^2 + 16 J^2} - \frac{16}{5} \frac{\tilde{J}^2}{\gamma_0^2} + \mathcal{O} \left(\frac{\tilde{J}^4}{ \gamma_0^4}\right) + \mathcal{O} \left(\frac{\tilde{J}^2}{ \gamma_0^2 } \frac{\gamma_0^3}{J^3}\right).
\end{equation}
The error scales linearly with the bandwidth $\mathcal{B} = \tau^{-1}$ of the photon, $\tilde{J}^2/\gamma_0^2 \propto \tau^{-1} \gamma_0^{-1} \propto \mathcal{B} \gamma_0^{-1}$ (faster than for the case of a Gaussian photon).

\section{V. Robustness to imperfections}

In this section, we assess the robustness of the entangled photonic state generation protocol against two sources of errors. First, independent photon emission into modes outside the half-waveguide at a rate $\gamma'=\epsilon \gamma_0$, modeled by the Lindbladian $\sum_{n=1}^3 \gamma ' \left(\hat{\sigma}_n \hat{\rho} \hat{\sigma}_n^\dagger - \frac{1}{2} \{\hat{\sigma}_n^\dagger \hat{\sigma}_n, \hat{\rho} \} \right)$. Second, positional errors in the emitter arrangement, which alter the coherent and dissipative couplings between the emitters. This, in turn, modifies the collective level structure, leading to imperfections in gate implementation. We analyze this in two ways: (i) assuming a modified spacing $d = \lambda_0 + \epsilon \lambda_0$, where $\epsilon \lambda_0$ represents the additional displacement; and (ii) introducing random position disorder following a Gaussian distribution with standard deviation $\sigma = \epsilon \lambda_0$, such that each emitter's position is given by $x_n = s + nd + \mathcal{N}(0,\sigma^2)$. In all cases, $\epsilon$ represents the magnitude of the perturbation with respect to the relevant scales.

\begin{figure}
    \includegraphics[width=\textwidth]{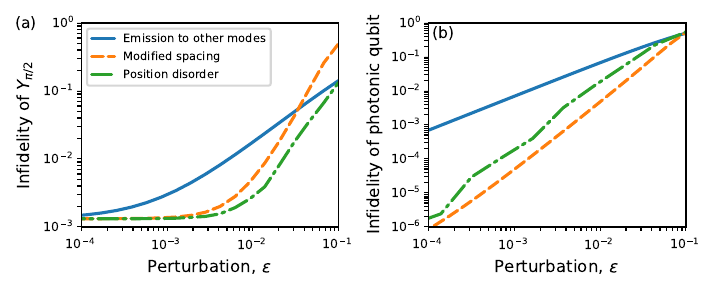}
    \caption{Gate infidelity $1 - \mathcal{F}$ for two sources of error. First, independent photon emission into modes outside the half-waveguide at a rate $\gamma'=\epsilon \gamma_0$ (blue traces). Second, positional errors in the emitter arrangement, arising from a modified spacing $d = \lambda_0 + \epsilon \lambda_0$ (dashed orange traces) and random position disorder such that $x_n = s + nd + \mathcal{N}(0,\epsilon^2 \lambda_0^2)$ (dash-dotted green traces). (a) Infidelity for the $Y_{\pi/2}$ gate between $|G\rangle$ and $|D\rangle$. (b) Infidelity of the emitted photonic wavepacket for a target Gaussian wavepacket of width $\tau = 7/(4\gamma_0)$. The infidelity arising from random position disorder is an average over $100$ realizations in (a) and $50$ realizations in (b).}
    \label{fig: errors_SM}
\end{figure}

To evaluate the robustness of the gates within the decoherence-free subspace (DFS), we focus on the $Y_{\pi/2}$ gate between $|G\rangle$ and $|D\rangle$. We numerically obtain the unitary evolution $\tilde{U}$ implemented by the system's non-Hermitian Hamiltonian, incorporating either decay to other modes or modified coherent and dissipative couplings due to positional shifts. The gate fidelity is then computed as in Section II. As shown in Fig.~\ref{fig: errors_SM}(a), the infidelity remains constant for small perturbations $\epsilon$, where it is set by the coupling to two-excitation states induced by the driving field (as discussed in Section II). For larger perturbations, their impact dominates and leads to an increase in infidelity.

We also assess the fidelity of the photon emission process. Given a target photon wavepacket $\psi_\mathrm{target}(t)$ and the wavepacket of the emitted photon under the optimal coupling from Eq.~(\ref{eq: optimal_coupling}), $\psi_\mathrm{out}(t)$, we define the fidelity as $\mathcal{F} = | \int_0^{T_\mathrm{em}} dt \psi_\mathrm{target}^*(t) \psi_\mathrm{out}(t)|^2 $. As discussed in Section III, the wavepacket of the emitted photon in the ideal configuration ($x_n=(n+1/4)\lambda_0$) is $\psi_\mathrm{out} (t)= \sqrt{3 \gamma_0} b(t) = \sqrt{\gamma_0} \sum_{n=1}^3 \langle \sigma_n \rangle(t)$. For general configurations, the photon wavepacket is $\psi_\mathrm{out} (t)= \sum_{n=1}^3 \sqrt{\Gamma_{nn}}  \langle \hat{\sigma}_n \rangle(t) = \sqrt{\gamma_0} \sum_{n=1}^3 \sin (k_0 x_n )\langle \hat{\sigma}_n \rangle(t)$, and can be obtained by solving the dynamics of the system within the single-excitation subspace under the non-Hermitian Hamiltonian. In Fig.~\ref{fig: errors_SM}(b), we show that the infidelity increases with $\epsilon$ for a Gaussian photon given by Eq.~(\ref{eq: SI_gaussian_adiabatic}).

\section{VI. Photonic entangled state generation with two two-level emitters coupled to a half-waveguide}

The conditional photon emission (CPE) gate for a collective system composed of three two-level emitters relies on the existence of the auxiliary dark state $\ket{A}$, whose coupling to the bright state $\ket{B}$ can be controlled over time (see Section IV of the Supplemental Material). This provides two main advantages. First, it allows to emit a single photon using only classical driving fields, such that the states of photonic qubit correspond to zero or one photon exactly. Second, it allows for an efficient and simple control of the wavepacket of the photonic qubit.

Photonic entangled state generation is also possible with just two two-level emitters coupled to the half-waveguide at distances $x_n = (n+1/4) \lambda_0$ from the mirror. The decoherence-free subspace is now formed by two states, the ground state $\ket{G} \equiv \ket{gg}$ and the single-excitation dark state $\ket{D} \equiv (\ket{eg} - \ket{ge})/\sqrt{2}$, which also form the matter qubit. The remaining two states, $\ket{B} \equiv (\ket{eg} + \ket{ge})/\sqrt{2} $ and $\ket{E} \equiv \ket{ee}$ are bright and decay at a rate $2\gamma_0$. Again, arbitrary rotations within the decoherence-free subspace are obtained via a weak classical field of the form $\hat{H}_{\Omega,D} = \Omega (\hat{\sigma}_1 - \hat{\sigma}_2)/\sqrt{2} + h.c.$, such that population of bright states is prevented by the Zeno effect ($\Omega \ll \gamma_0$) and the photon blockade (for large coherent exchange interactions $J \gg \Omega_0$ between emitters).

Due to the absence of an auxiliary dark state, the CPE gate needs to be performed by rapidly transferring the amplitude in $\ket{G}$ to a bright state via the drive $\hat{H}_{\Omega,B} = \Omega_B (\hat{\sigma}_1 + \hat{\sigma}_2)/\sqrt{2} + h.c. = \Omega_B \hat{\sigma}_B + h.c.$ in a time $T \ll \gamma_0^{-1}$ (such that no photon emission occurs during the transfer). In the absence of a coherent exchange interaction $J$ between the emitters, $\ket{G} \leftrightarrow \ket{B} \leftrightarrow \ket{E}$ are resonantly coupled, and only the doubly-excited state $\ket{E}$ can be prepared after a time $T = \pi / \sqrt{2} \Omega$. As a result, the two states of the photonic qubit correspond to zero and two photons. Generation of photonic qubits encoded as zero and one photons requires a non-linearity that allows the transfer $\ket{G} \rightarrow \ket{B}$ while suppressing the coupling between $\ket{B}$ and $\ket{E}$. For superconducting qubits, this can be achieved by applying $\hat{H}_{\Omega,B}$ for a time $T = \pi /2 \Omega$ under a large coherent interaction $J \gg \Omega \gg \gamma_0$ that renders both transitions off-resonant by $2 J$. For neutral atoms coupled to waveguides, the non-linearity can be attained via Rydberg interactions if both atoms are within a Rydberg blockade radius. Additionally, control of the wavepacket of the emitted photons requires a tunable coupling strength $\gamma_0$ of the emitters to the waveguide over time~\cite{Painter_states}. This stays in stark contrast to the three-emitter setup, where the auxiliary dark state allows to control the wavepacket by simply varying the detuning of one emitter over time.  

Finally, the controlled phase gate or CZ gate can be easily achieved in the case of two emitters coupled to a waveguide. The reflection coefficient for an incoming photon in the low intensity limit reads $r = 1 - 2 i \gamma_0 \langle \hat{\sigma}_B \rangle / \Omega_B$. Since the single-excitation dark state $\ket{D}$ is decoupled from the electromagnetic field of the waveguide, an incoming photon does not interact with $\ket{D}$ ($\langle \hat{\sigma}_B \rangle = 0$), leading to $r_d = 1$. If the emitter system is in $\ket{G}$, however, the amplitude in the bright state in the low intensity limit is $b(t) \approx -  \Omega / ( \Delta - i \gamma_0)$. Noting that $\langle \hat{\sigma}_B \rangle = b$, we obtain the reflection coefficient for the ground state $r_g = 1 - 2 \gamma_0 / (\gamma_0 +  i \Delta)$. Thus, the scattered photon acquires a phase flip ($r_g = -1$) if it is on resonance with the $\ket{G} \leftrightarrow \ket{B}$ transition ($\Delta = 0$), thereby implementing the CZ gate given in Eq.~(\ref{eq: CZ_gate_ideal}).

\end{document}